\documentclass{optica-article}

\journal{opticajournal} 

\articletype{Research Article}

\usepackage{lineno}
\usepackage{physics}

\begin{document}

\title{Quantized topological phases beyond square lattices in Floquet synthetic dimensions}

\author{Samarth Sriram,\authormark{1} Sashank Kaushik Sridhar,\authormark{2} and Avik Dutt\authormark{2,3,4, *}}

\address{\authormark{1}Department of Physics, University of Maryland, College Park, MD 20742, USA\\
\authormark{2}Department of Mechanical Engineering, University of Maryland, College Park, MD 20742, USA\\
\authormark{3}Institute for Physical Science and Technology, University of Maryland, College Park, MD 20742, USA

\authormark{4}National Quantum Laboratory (QLab) at Maryland, College Park, MD 20740, USA}

\email{\authormark{*}avikdutt@umd.edu} 


\begin{abstract}
Topological effects manifest in a variety of lattice geometries. While square lattices, due to their simplicity, have been used for models supporting nontrivial topology, several exotic topological phenomena such as Dirac points, Weyl points and Haldane phases are most commonly supported by non-square lattices. Examples of prototypical non-square lattices include the honeycomb lattice of graphene and 2D materials, and the Kagome lattice, both of which break fundamental symmetries and can exhibit quantized transport, especially when long-range hoppings and gauge fields are incorporated. The challenge of controllably realizing such long-range hoppings and gauge fields has motivated a large body of research focused on harnessing lattices encoded in "synthetic" dimensions. Photons in particular have many internal degrees of freedom and hence show promise for implementing these synthetic dimensions; however, most photonic synthetic dimensions has hitherto created 1D or 2D square lattices. Here we show that non-square lattice Hamiltonians such as the Haldane model and its variations can be implemented using Floquet synthetic dimensions. Our construction uses dynamically modulated ring resonators and provides the capacity for direct $k$-space engineering of lattice Hamiltonians. This $k-$space construction lifts constraints on the orthogonality of lattice vectors that make square geometries simpler to implement in lattice-space constructions, and instead transfers the complexity to the engineering of tailored, complex Floquet drive signals. We simulate topological signatures of the Haldane and the brick-wall Haldane model and observe them to be robust in the presence of external optical drive and photon loss, and discuss unique characteristics of their topological transport when implemented on these Floquet lattices. Our proposal demonstrates the potential of driven-dissipative Floquet synthetic dimensions as a new architecture for $k$-space Hamiltonian simulation of high-dimensional lattice geometries, supported by scalable photonic integration, that lifts the constraints of several existing platforms for topological photonics and synthetic dimensions.  
\end{abstract}

\section{\label{sec:level2}Introduction} 
The quantum Hall effect, first discovered by von Klitzing and collaborators in 1980 \cite{von-kltizing}, paved the way for a new class of materials with properties defined not by local symmetry but by global topology, eventually establishing the field of topological insulators in the mid-2000s \cite{Bernavig-Hughes-Zhang, Bernavig-Hughes, Qi-Hughes-Zhang, Fu-Kane-Mele, Kane-Mele}. These materials showed hallmark features of quantized, unidirectional transport, and also set new standards for the measurement of resistance \cite{von_klitzing_metrology_2017}. Traditionally, quantum Hall phases required a strong external magnetic field, until Haldane in his seminal paper \cite{HaldaneOGpaper} showed that the essence of these topological phases lies in the breaking of time reversal symmetry, which can equivalently be done by incorporating next-nearest neighbor hopping. This effect was later dubbed the quantum anomalous Hall effect (QAHE) \cite{chang_experimental_2013}. Haldane’s discovery marks a characteristic telltale sign for systems exhibiting topological phases, i.e.,  non-square nature of lattices, making the Haldane model an important and sought-after discovery. While experiments involving the Haldane model are challenging due to the necessity of strong control over next-nearest-neighbor tunneling, various innovative approaches in ultra-cold atoms and Fe-based ferromagnetic insulators \cite{HaldExp1, HaldExp3} have been proposed. However, the Haldane model in its original form has so far only been realized experimentally using ultracold fermions \cite{HaldExp2} and light-driven graphene \cite{mciver_light-induced_2020}. 

To gain versatile control in experiments, the concept of synthetic dimensions has emerged, which in contrast to real-space,  are formed by the coupling of internal degrees of freedom like spin, angular momentum or frequency \cite{Yuan:18, tpht4, boada_quantum_2012, celi_synthetic_2014, ozawa_topological_2019-1}. Introducing such couplings allows for the probing of higher dimensional physics in systems of lower dimensions. Photonics in particular has been a successful platform to explore synthetic dimensions due to the scalability it offers, and due to the reduction of spatial complexities associated with system design into tasks in signal processing and control \cite{price_roadmap_2022}. A prime focus of research in photonic synthetic dimensions is the study of topological physics such as Hermitian and non-Hermitian effects, gauge fields \cite{fang_realizing_2012}, and higher order topological insulators \cite{yuan_synthetic_2021, lustig_photonic_2019, leefmans_topological_2022, chalabi_synthetic_2019, dutt_higher-order_2020, Yuan:16, yuan_creating_2020, dutt_single_2020, dutt_experimental_2019-1, ozawa_synthetic_dim_2016, wang_generating_2021, wang_topological_2021, weidemann_topological_2020}.

Despite significant progress, most experimental work in synthetic dimensions has realized 1D lattices or 2D square lattices. In photonics, this includes work in frequency modes \cite{yuan_synthetic_2021, senanian_programmable_2023, dutt_experimental_2019-1, dutt_single_2020, wang_generating_2021, wang_topological_2021, cheng_artificial_2023, qin_spectrum_2018, balcytis_synthetic_2022, suh_photonic_2024, dinh_reconfigurable_2024, englebert_bloch_2023}, temporal modes \cite{chalabi_synthetic_2019, leefmans_topological_2022, leefmans_topological_2024, wimmer_experimental_2017}, orbital angular momentum modes \cite{luo_quantum_2015, luo_topological_2018, yang_realization_2023}, waveguide supermodes \cite{lustig_photonic_2019} and photon-number lattices \cite{saugmann_fock-state-lattice_2023, biswas_topological_2024, deng_observing_2022, cheng_arbitrary_2021}. However, several important topological models exist on non-square lattices -- the hexagonal or honeycomb lattice being a prime example. The hexagonal lattice hosts graphene's Dirac points, and with the addition of next-nearest-neighbor (NNN) complex-valued coupling hosts Haldane phases as mentioned earlier \cite{HaldaneOGpaper, HaldExp2}. In atomic systems that harness internal degrees of freedom \cite{PriestleyBrickWall, fangzhao_direct_2017,oliver_bloch_2023,sundar_synthetic_2018}, recent theoretical efforts \cite{Agrawal_twodimensional_2024} have shown prospects for achieving the Haldane model and other general 2D constructions. While real-space photonic lattices have shown significant success in realizing hexagonal geometries \cite{rechtsman_photonic_2013, noh_topological_2018, cerjan_experimental_2019}, complex-valued NNN coupling to realize Haldane models is challenging and has only been proposed \cite{minkov_haldane_2016} and demonstrated recently \cite{liu_gain-induced_2021}. The challenges in synthetic dimensions are the opposite: complex-valued NNN coupling is straightforward \cite{dutt_experimental_2019-1, wang_multidimensional_2020, bell_spectral_2017} but non-square (e.g. hexagonal) lattices are difficult as the lattice vectors are no longer independent and orthogonal. Intriguing protocols towards this direction have been proposed \cite{Luqiyuan}, but they require several rings with precisely shifted resonance frequencies to produce modified versions of the original Haldane model. Hence, the question arises if the original Haldane model, and more generally, non-square lattices, can be created using photonic synthetic dimensions. 

In this paper, we explicitly construct the original Haldane model in synthetic dimensions, and show a path to realize non-square lattices using a modulated photonic molecule. This proposal uses our recently introduced concept of Floquet synthetic dimensions \cite{SashankPaper}, which allows for direct $k$-space engineering of two-band Hamiltonians. We numerically demonstrate quantized topological pumping, which is a hallmark of the quantum anomalous Hall effect, in both the Haldane model and a simplified version called the brick-wall Haldane model \cite{PriestleyBrickWall}. Importantly, the topological pumping survives with optical and microwave drives and is robust to dissipation, making the proposed construction experimentally feasible with current photonic technology. We also discuss unique subtleties in the experimental protocols that need to be taken into consideration due to the non-orthogonal nature of the the lattice vectors in these hexagonal models. It is anticipated that our work will substantially increase the classes of Hamiltonians that can be created in high-dimensional lattices beyond the square, cubic and hypercubic geometries. 

\section{\label{sec:level5}The Haldane and brick-wall Haldane Models}
Throughout this paper, we use two-band Hamiltonians that exhibit non-trivial topology. We first briefly review two topological Hamiltonians that are characterized by complex non-square geometries: the Haldane \cite{HaldaneOGpaper} and brick-wall Haldane \cite{PriestleyBrickWall} models. As materials characterized by these Hamiltonians break time-reversal-symmetry, particle-hole symmetry and chiral symmetry, they belong to the A (unitary) class of topological insulators in the tenfold way \cite{ryu_topological_2010}.
\begin{figure}
        \centering
        \includegraphics[scale=0.35]{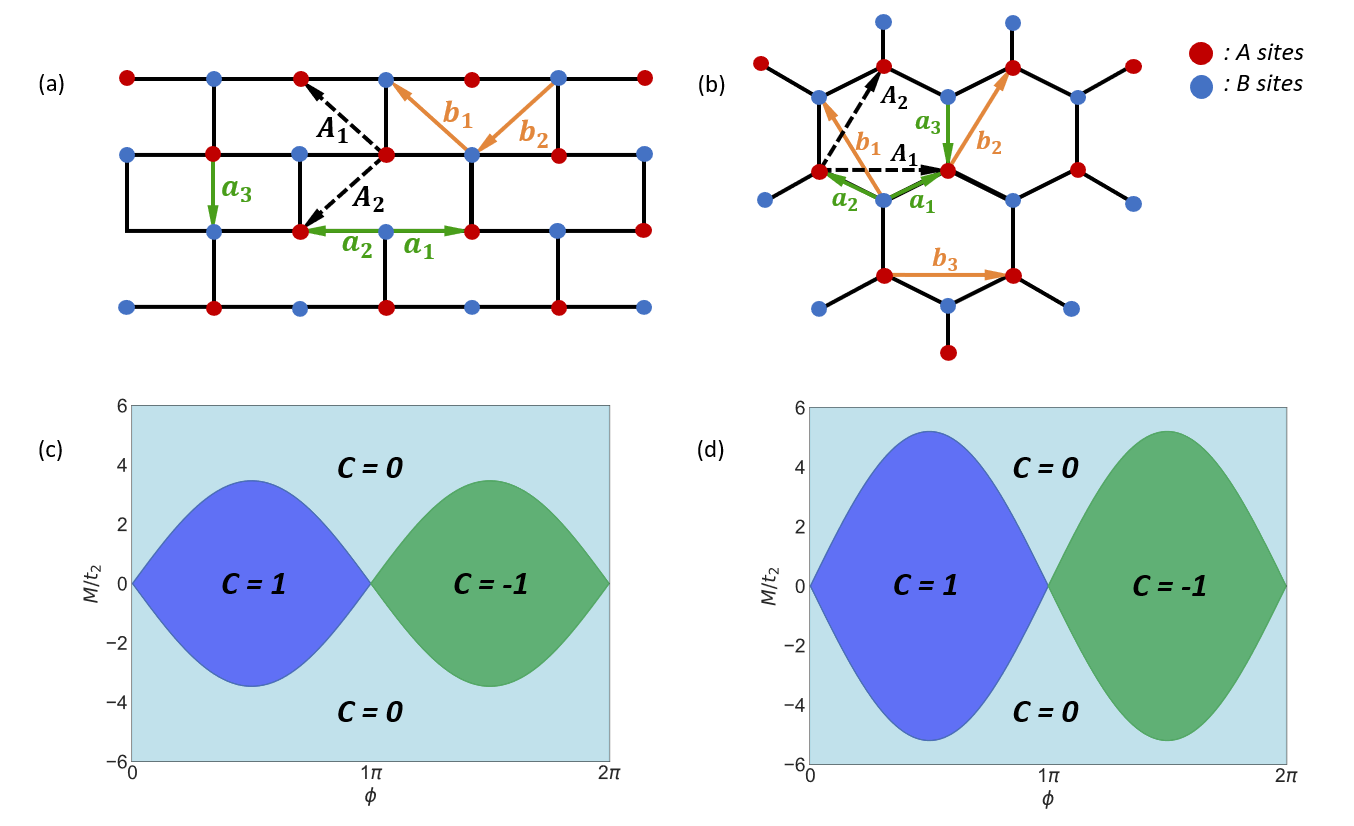}
        \caption{\textbf{(a)} Real-space lattice for brick-wall Haldane model. The brick-wall lattice is obtained by taking the top and bottom sites of the graphene lattice and moving them down. Nearest-neighbor (NN) hopping vectors are given by $\vec{a}_1 = (1, 0)$, $\vec{a}_2 = (-1, 0)$ and $\vec{a}_3 = (0, -1)$. Next-nearest-neighbor (NNN) hopping vectors are given by $\vec{b}_1 = \vec{a}_2 - \vec{a}_3$ and $\vec{b}_2 = \vec{a}_3 - \vec{a}_1$. The geometry of the space removes NNN hopping of $\vec{b}_3 = \vec{a}_1 - \vec{a}_2$. \textbf{(b)} Real-space lattice for the Haldane model. NN hopping vectors are given by $\vec{a}_1 = 1/2(\sqrt{3}, 1)$, $\vec{a}_2 = 1/2(-\sqrt{3}, 1)$ and $\vec{a}_3 = (0, -1)$, while NNN hopping vectors are given by $\vec{b}_1 = \vec{a}_2 - \vec{a}_3$, $\vec{b}_2 = \vec{a}_3 - \vec{a}_1$ and $\vec{b}_3 = \vec{a}_1 - \vec{a}_2$. For both (a) and (b), "A" subsites are indicated by red while "B" subsites are indicated by blue. Phase diagrams for \textbf{(c)} the brick-wall Haldane and \textbf{(d)} Haldane models. Phase transitions between trivial and topological phases occur for $M/t_2 = \pm 2\sqrt{3}\sin(\phi)$ for brick-wall Haldane and at $M/t_2 = \pm 3\sqrt{3}\sin(\phi)$ for Haldane models, respectively. The topological phases are characterized by non zero Chern number $C = \pm 1$, while the trivial phase is characterized by $C = 0$. In (a) and (b), lattice vectors are marked by dashed lines and are given by $\vec{A}_1$ and $\vec{A}_2$. 
        }
        \label{fig:HaldBWHAllPlots}
\end{figure}

In 1988, Haldane \cite{HaldaneOGpaper} theoretically demonstrated the presence of quantized currents without the need for external magnetic fields. By taking a tight-binding model on a honeycomb lattice composed of two sets of sites (labeled "A" and "B," Fig. \ref{fig:HaldBWHAllPlots}), one can show that edge currents can be generated by the combined breaking of inversion and time-reversal symmetry (TRS), rather than the presence of a net non-zero magnetic flux. This phenomenon is now well understood and is referred to as the quantum anomalous Hall effect (QAHE) \cite{chang_experimental_2013}. The tight-binding Hamiltonian is given by
\begin{equation}
    H = -t_1\sum_{\langle i, j \rangle}c_i^\dag c_j - t_2\sum_{\langle \langle i, j \rangle \rangle}e^{-i\phi_{ij}}c_i^\dag c_j + \sum_i t_i c_i^\dag c_i.
    \label{TBHam}
\end{equation}
Here, $c_i$ ($c_i^\dag$) represent annihilation (creation) operators for $i \in \{A_n, B_n\}$, where $n$ indexes the unit cell. The first and second terms represent the nearest-neighbor (NN) and next-nearest-neighbor (NNN) hopping, with strengths $t_1$ and $t_2$ respectively. A complex phase, $\phi_{ij}$, is included in the NNN term and is responsible for breaking the TRS. The last term represents the on-site mass, where $t_i = M$ for $i = A_n$, and $t_i = -M$ for $i = B_n$, and is responsible for breaking the inversion symmetry. The above Hamiltonian can be represented in Bloch form (or momentum space) for the case of periodic boundary conditions:
\begin{equation}
    H = \sum_{\mathbf{k}}\mathbf{\Psi}_{\mathbf{k}}^\dag \mathcal{H}(\mathbf{k})\mathbf{\Psi}_{\mathbf{k}}
\end{equation}
where $\Psi_\mathbf{k} = (\psi_{Ak}, \psi_{Bk})^T$ is the Bloch spinor, and $\mathcal{H}(\mathbf{k})$ is the Bloch Hamiltonian
\begin{eqnarray}
    \mathcal{H}(\mathbf{k}) &= 2t_2 \cos(\phi)\sum_i\cos(\mathbf{k} \cdot \mathbf{b}_i)\mathbf{I} + \, t_1\sum_i\left[ \cos(\mathbf{k}\cdot\mathbf{a}_i)\sigma_x + \sin(\mathbf{k}\cdot\mathbf{a}_i)\sigma_y\right] + \, \nonumber
    \\&\left[ M - t_2 \sin(\phi) \sum_i \sin(\mathbf{k}\cdot\mathbf{b}_i) \right]\sigma_z.
\label{eq:HaldaneBlochForm}
\end{eqnarray}
In this Hamiltonian, $\mathbf{a}_i$ and $\mathbf{b}_i$ represent the directions of NN and NNN hopping, respectively. $\sigma_i$ represent the 2$\times$2 Pauli matrices for $i\in \{x, y, z\}$, and $\mathbf{I}$ is the 2$\times 2$ identity matrix. 
The above Hamiltonian can be re-written as
\begin{equation}
    \mathcal{H}(\mathbf{k}) = \epsilon(\mathbf{k})\mathbf{I} + \mathbf{d}(\mathbf{k}) \cdot \vec{\mathbf{\sigma}}
    \label{eq:shortbloch}
\end{equation}
Following \cite{bernevig_topological_2013, PriestleyBrickWall}, for the case of periodic boundary conditions, it can be shown that the Haldane model is characterized by a topological invariant called the Chern number, which can take on three values $C=-1, 0, 1$ (Fig. \ref{fig:HaldBWHAllPlots}(d)). By introducing physical boundaries to the system, the conventional bulk-boundary correspondence gives us the quantized Hall conductance $\sigma_{xy} = Ce^2/h$ despite the absence of net non-zero flux.

Recently, Priestley et. al. \cite{PriestleyBrickWall} showed that a simplified Haldane-like model on a brick-wall lattice can be realized in the synthetic dimensions of atoms. By adiabatically "straightening out" the honeycomb, the hexagonal lattice can be transformed into the brick-wall lattice (Fig. \ref{fig:HaldBWHAllPlots}(a, b)). Equivalently, such a transformation is achieved by moving down the top and bottom sites of the hexagon. The consequent changes to the model are the NN and NNN vector directions, i.e., $\mathbf{a}_i$ and $\mathbf{b}_i$ respectively (see Fig. \ref{fig:HaldBWHAllPlots} captions). Geometric arguments as well as explicit calculations reveal that the two models are topologically equivalent, since the topological invariant -- the Chern number -- remains unchanged within the boundaries of the distinct topological phases, although the location of the topological-to-trivial phase transition boundaries changes (Fig \ref{fig:HaldBWHAllPlots}(c) and \ref{fig:HaldBWHAllPlots}(d)) \cite{PriestleyBrickWall}.

The topological-to-trivial phase transitions occurs at the points in phase space constituting the opening and closing of the band gaps. These points occur when $|\mathbf{d}(\textbf{k})| = 0$, which translates to phase boundaries at (Figs. \ref{fig:HaldBWHAllPlots}(c,d))
\begin{equation}
\frac{M}{t_2} =
\left\{
\begin{array}{ll}
    \pm 3\sqrt{3}\sin(\phi), & \text{Haldane} \\
    \pm 2\sqrt{3}\sin(\phi), & \text{brick-wall Haldane}
\end{array}
\right.
\label{eq:phaseboundaryvals}
\end{equation}
As these models are respectively characterized by a hexagonal lattice and a rectangular lattice with alternating couplings. Henceforth, we refer to both these models as being on non-square lattices, especially since they are topologically equivalent.

{Although originally viewed as real-space models, the crux of the model lies in the topology of the system, allowing us to extrapolate these models to other channels of transport; herein lies the utility of synthetic dimensions. We therefore consider the temporal analogue of the Bloch hamiltonian - the Floquet hamiltonian - which will allow us to simplify the physical implementation to a single spin-1/2 particle.}

\section{\label{sec:level3}The Floquet Lattice}
\begin{figure}
    \hspace{-.4in}
    \includegraphics[scale=0.2]{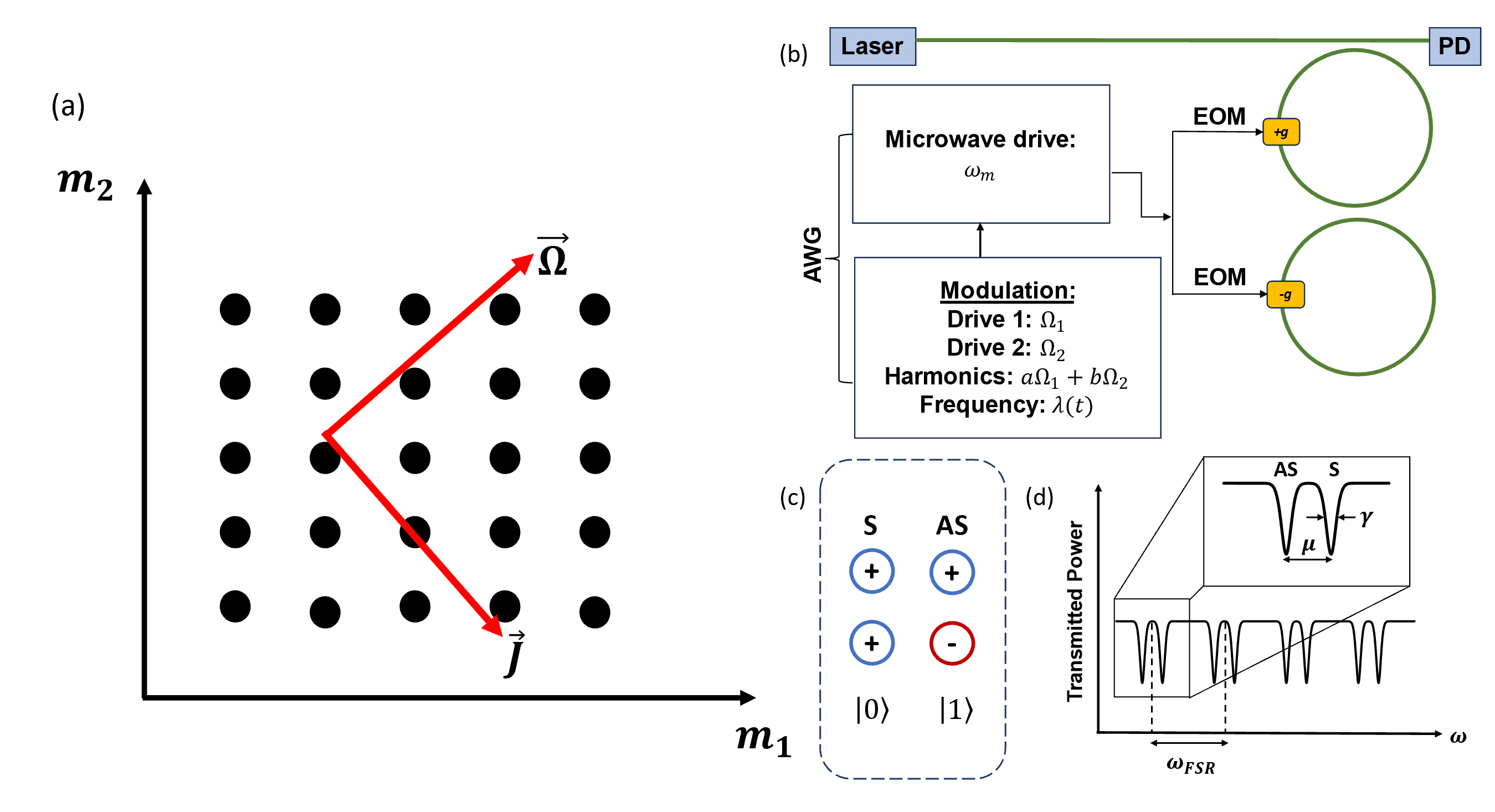}
    \caption{(a) Pictorial representation of a 2D Floquet-lattice. The lattice responds to an external force/electric field ($\vec{\Omega}$) by an orthogonal current ($\vec{J}$). Each point on the lattice is a harmonic of the drives. More generally, $n$ external drives of mutually incommensurate frequencies produces an $n$-dimensional lattice in Floquet synthetic dimensions. 
    (b) Schematic setup for the photonic molecule. To simulate non-square lattices, we need extra harmonics of the form $a\Omega_1 + b\Omega_2$, where $a$ and $b$ $\in \mathbb{R}$. These harmonics also lead to long-range couplings (as in the Haldane model). (c) Schematic representation of the eigenmodes, labeled Symmetric (S) and Anti Symmetric (AS). 
    (d) Transmission spectra of the S and AS modes. These modes are separated by `$\mu$'. The coupling rate with the bus waveguide is given by $\gamma_e$ and the loss associated with each resonator is $\gamma$.}
    \label{fig:floq_latt}
\end{figure}
In this section, we briefly review the concept of Floquet synthetic dimensions \cite{SashankPaper} based on multiple incommensurate frequency drives as introduced by Martin, Refael and Halperin \cite{MRH}.

The physics of a quantum system evolving under a time-dependent Hamiltonian is captured by the Schr\"{o}dinger equation 
\begin{equation}
    i\frac{\partial}{\partial t}\ket{\psi(t)} = \mathcal{H}(t) \ket{\psi(t)}
    \label{eq:tdse}
\end{equation}
where $\mathcal{H}(t)$ is the time-dependent Hamiltonian and $\ket{\psi(t)}$ is the quantum state that describes the system. 
In analogy to spatially periodic Hamiltonians, which are diagonalizable using Bloch's theorem, Floquet's theorem lays the basis for temporally periodic systems. In the latter, we replace the quasi-momentum $k$ of the particles with their quasi-energy $E$. In Bloch's theorem, the eigenstates (Bloch states) of the system are conveniently found by diagonalizing the translation operator. In Floquet's theorem, the eigenstates (Floquet states) are given by diagonalizing the single period time evolution operator. 

Throughout this paper, we consider the case of a multi-tone drive described by a quasi-periodic Hamiltonian \cite{david_long} $\mathcal{H}(t) = \mathcal{H}(\vec{\theta}(t)) = \mathcal{H}(\vec{\Omega}t + \vec{\theta}_0)$, where $\vec{\Omega}$ and $\vec{\theta}_0$ describe the frequency and initial phase of each drive. It is easy to observe that $\vec{\theta} \in \mathbb{T}^n$ and spans the "Floquet-zone," in analogy to the Brillouin-zone for the case of spatially periodic systems.

By expanding  the Hamiltonian and periodic part of the Floquet states in a Fourier series,  and plugging them into Schr\"{o}dinger's equation, we get the Floquet representation of the Hamiltonian

\begin{equation}
    \small
    E_\mu\ket{\phi_{\mu, \vec{n}}} = \sum_{\vec{m} \, \in \, \mathbb{Z}^{n}}\mathcal{R}_{\vec{n}, \vec{m}}\ket{\phi_{\mu, \vec{m}}}
    \label{eq:floquet_tb}
\end{equation}
where 
\begin{equation}
    \mathcal{R}_{\vec{n}, \vec{m}} = H_{\vec{n}-\vec{m}}e^{-i(\vec{n}-\vec{m})\cdot\vec{\theta}_0} - \vec{n}\cdot\vec{\Omega}\delta_{\vec{n}\vec{m}}.
\end{equation}
Where $E_\mu$ are the quasi-energies, and $\ket{\phi_\mu}$ are the Floquet eigenstates. The above equation mimics a tight binding model on a photon number lattice, also known as the Floquet lattice. A pictorial representation of a square Floquet lattice is given by Fig. \ref{fig:floq_latt}. Such a  lattice serves as the basis for a mapping between a $d$-dimensional system, which is being driven by $n$ external drives of mutually incommensurate frequencies, and a static model, which is represented using the tight binding lattice, with $n$ synthetic dimensions. In addition, one can observe an on-site potential term, $U(\vec{m}) = -\vec{m} \cdot \vec{\Omega}$, that corresponds to an applied external electric field $\mathcal{E} = \vec{\Omega}$. For incommensurately related frequencies, this term is non-zero $\forall \, \vec{m} \in \mathbb{Z}^n$. So, in the presence of non-zero Berry curvature, the system responds by an orthogonal current, with anomalous velocity
\begin{equation}
    \vec{v} = \vec{\Omega} \times \mathcal{F}_{\vec{\theta}}
\end{equation}
where $\mathcal{F}_{\vec{\theta}}$ is the Berry-curvature. The Floquet lattice representation for the Haldane and brick-wall Haldane model Hamiltonians for the general lattice configuration is

\begin{eqnarray}
     \nonumber i \partial_t \psi_{\vec{n}} &=& \frac{1}{2} \sum_{j}  \sigma_+ e^{i \vec{\theta}_0 \cdot \vec{a}_j} \psi_{\vec{n} - \vec{a}_j} + \sigma_- e^{-i \vec{\theta}_0 \cdot \vec{a}_j} \psi_{\vec{n} + \vec{a}_j} \\
     \nonumber && + \, i \, \sin(\phi) \sigma_z \left( e^{i \vec{\theta}_0 \cdot \vec{b}_j} \psi_{\vec{n} - \vec{b}_j} - e^{-i \vec{\theta}_0 \cdot \vec{b}_j} \psi_{\vec{n} + \vec{b}_j} \right) \\ 
     && + \left(\delta\sigma_z - \vec{n}\cdot\vec{\Omega}\right),
\end{eqnarray}
where $j\in \left\{1, 2, 3\right\}$,  $\vec{a}_j$ and $\vec{b}_j$ are the nearest-neighbor and next-nearest-neighbor hopping respectively, and $\sigma_{\pm} = \sigma_x \mp i\sigma_y$. In contrast to square lattices, $\vec{n}$ is written in the basis of non-orthogonal lattice vectors, that is, the lattice vectors spanning the graphene lattice. 

One can observe that the incommensurate nature of the drives is a necessary  condition to observe phenomena like quantized pumping. In this case, the system can be considered quasi-periodic, as the dynamics cannot repeat itself. For the case of commensurate frequencies it is easy to see, $\exists \, \vec{l} \, \in \mathbb{Z}^n \ni$ $\vec{l} \cdot \vec{\Omega} = 0 \implies \forall \vec{m} \in \mathbb{Z}^n$, $\vec{m} \cdot \vec{\Omega} = (\vec{l} + \vec{m}) \cdot \vec{\Omega}$. Effectively, the points $\vec{m}$ and $\vec{m} + \vec{l}$ are topologically equivalent, and the Floquet lattice collapses into a cylinder. In Sec. 5.2 and Fig. \ref{fig:allmodesplot}, we observe this difference between dynamics driven by commensurate (rational $\Omega_1/\Omega_2$) and incommensurate (irrational $\Omega_1/\Omega_2$) ratios of the frequencies. Our simulations reveal that although the pumping behavior changes when the mass term $M$ is changed from the topological to the trivial regime (Eq. \eqref{eq:phaseboundaryvals}), no equal and opposite quantized pumping consistent with the Chern number is observed in the topological phase, and the sharp transition at the boundaries of the phase diagram is lost.

While it is possible in principle to realize this Floquet Hamiltonian using a spin-1/2 particle driven by incommensurate harmonics in orthogonal directions, the adiabatic dynamics are often experimentally challenging to implement without counter-diabatic drives \cite{delcampo_shortcuts_2013} within the coherence lifetimes of qubit platforms \cite{malz_topological_2021, boyers_exploring_2020}. The non-square lattice implementation also exacerbates the spatial complexity of the experiment due to the need for additional harmonics along each axis. To address these issues, we consider a driven-dissipative photonic molecule, as introduced in Ref. \cite{SashankPaper}. With a photonic molecule comprised of coupled-ring resonators, one can realize such a Floquet lattice at timescales much longer than the photon lifetimes of each ring.

\section{\label{sec:level4}Realizing a Floquet Lattice: photonic molecule}

The term `photonic molecule' refers to pairs or arrays of coupled optical resonators. Initially studied in analogy with atomic systems \cite{PMopticalModes1, PMopticalModes2}, they have since become a valuable resource for engineering driven-dissipative dynamics \cite{DriveDissPM1, DriveDissPM2}, Fano resonances \cite{FanoResonancePM1} and ultrafast optical switching \cite{UltrafastSwitchingPM1}. Studies with dynamically-modulated photonic molecules have demonstrated utility for optical memories \cite{PMopticalModes3SFANEP} and RF-optical conversion \cite{singh_coupled_2024}, and can also be leveraged for emulating spin Hamiltonians \cite{PMopticalModes3SFANEP, SashankPaper}.

\subsection{\label{sec:level4.1}Construction of the photonic molecule}

The photonic molecule (Fig. \ref{fig:floq_latt}(b)) consists of two identically coupled cavities. The evanescent coupling between the resonant modes of the rings leads to the formation of non-degenerate supermodes, characterised as "symmetric" (S) and "anti-symmetric" (AS). Transitions between the supermodes can be controlled via the intra-cavity electro-optic modulators (EOMs) (see Figures \ref{fig:floq_latt}(c, d)), allowing us to treat each pair of supermodes as a spin-1/2 particle. 

In addition, by introducing adiabatic amplitude and phase modulation to the RF signal containing two or more incommensurate frequency tones, we can achieve two-band Hamiltonians encoded in the quasiperiodic temporal texture of the two eigen-energies.

\subsection{\label{sec:level3b} Hamiltonian of the photonic molecule}

Consider the photonic molecule with uncoupled bosonic modes $a_1$ and $a_2$ modulated by a signal $V(t)$. The Hamiltonian is given by
\begin{eqnarray}
    H &= \omega_0\left( a_1^\dag a_1 + a_2^\dag a_2 \right) + \frac{\mu}{2}\left(a_1^\dag a_2 + a_2^\dag a_1\right) \nonumber + gV(t)\left(a_1^\dag a_1 - a_2^\dag a_2\right).
\end{eqnarray}
where, the minus sign in the last term is implemented by a $\pi$ phase shift in the modulation between the two rings of the molecule, and $$V(t) = V_x(t)\cos(\omega_mt+\Delta(t)) + V_y(t)\sin(\omega_mt+\Delta(t)).$$ Here, $\omega_m = \mu-\delta$ is the RF carrier, $V_x(t)$ and $V_y(t)$ are the amplitude modulation and $\Delta(t)$ is the phase modulation. $g$ is the electro-optic transduction coefficient.

We now define the supermodes with the transformations: $b_1 := \frac{1}{\sqrt{2}}(a_1 + a_2)$ and $b_2 := \frac{1}{\sqrt{2}}(a_1 - a_2)$. This gives us the effective supermode Hamiltonian:
\begin{equation}
    H_{\text{SM}} = \omega_{+}b_1^\dag b_1 + \omega_{-}b_2^\dag b_2 + gV(t)\left(b_1^\dag b_2 + b_2^\dag b_1\right)
\end{equation}
where $\omega_{\pm}:=\omega_0 \pm \mu/{2}$. 
By centering the energy-axis at $\omega_0$, taking the interaction picture and subsequently making rotating wave approximations ($|gV_x(t)|$, $|gV_y(t)| \ll \omega_m$), we obtain \cite{SashankPaper}:  

\begin{equation}
    \mathcal{H}(t) = \frac{\delta - \lambda(t)}{2}\sigma_z + \frac{gV_x(t)}{2}\sigma_x + \frac{gV_y(t)}{2}\sigma_y
    \label{eq:floquetHamiltonian}
\end{equation}
where $\lambda(t)$ is the frequency modulation, defined as $\lambda := d/dt[\Delta(t)]$. Here we use the spin-1/2 operators $\sigma_{x,y,z}$ to represent the bosonic modes within the single excitation subspace as $\sigma_x = b_1^\dag b_2 + b_2^\dag b_1$, $\sigma_y -i(b_1^\dag b_2 - b_2^\dag b_1)$ and $\sigma_z = b_1^\dag b_1 - b_2^\dag b_2$.

Thus, we can achieve a generic time-dependent spin-1/2 Hamiltonian, which is a template for realizing two-band models in Floquet lattices. With appropriate amplitude and phase modulations, we will show how one can venture beyond \textit{square} lattices \cite{MRH, boyers_exploring_2020, long_nonadiabatic_2021, long_coupled_2022, SashankPaper} to realize the Haldane model in Eq. \eqref{eq:HaldaneBlochForm}.  
\section{\label{sec:level6}Temporal 
topological dynamics of the Haldane and brick-wall Haldane models}


To study topological Hamiltonians such as Eq. \eqref{eq:HaldaneBlochForm} on a Floquet lattice, we map the quasi-momentum $\mathbf{k}\equiv(k_x,k_y)$ to incommensurate frequencies, i.e., $k_x \rightarrow \theta_1(t)$ and $k_y \rightarrow \theta_2(t)$, where $\theta_1(t) = \Omega_1t + \phi_1$ and $\theta_2(t) = \Omega_2t + \phi_2$,\; $\Omega_1/\Omega_2 \neq p/q$ $\forall$ $p, q \in \mathbb{Z}$ and $q \neq 0$. For brevity, we represent $\vec{\theta} = \left(\theta_1(t), \theta_2(t)\right)$.


Comparing Eqs. \eqref{eq:HaldaneBlochForm} and \eqref{eq:floquetHamiltonian} while neglecting the $k$-dependent energy shift $\epsilon$($\mathbf{k}){\bf I}$, we get the exact expressions for the amplitude and phase modulation. Neglecing the diagonal part of the Hamiltonian does not change the eigenstates and hence leaves the topology of the model unchanged. For the Haldane model, the amplitude modulation is given by
\begin{eqnarray}
    \small
    \begin{array}{ll}
    V_x(t) &= V_0 \biggl\{\cos\left(\frac{\sqrt{3}}{2}\theta_1(t) + \frac{1}{2}\theta_2(t)\right) + \cos\left(-\frac{\sqrt{3}}{2}\theta_1(t) + \frac{1}{2}\theta_2(t)\right) + \cos\left(\theta_2(t)\right)\biggl\} \\
    
    V_y(t) &= V_0 \biggl\{\sin\left(\frac{\sqrt{3}}{2}\theta_1(t) + \frac{1}{2}\theta_2(t)\right) + \sin\left(-\frac{\sqrt{3}}{2}\theta_1(t) + \frac{1}{2}\theta_2(t)\right) - \sin\left(\theta_2(t)\right) \biggl\},
\end{array}
\label{eq:ampmodh}
\end{eqnarray}
while the frequency modulation is given by
\begin{equation}
\small
    \begin{array}{ll}
    \lambda(t) &= -2gV_0\sin(\phi)\biggl\{\sin\left(-\frac{\sqrt{3}}{2}\theta_1(t) + \frac{3}{2}\theta_2(t)\right) + \sin\left(-\frac{\sqrt{3}}{2}\theta_1(t)  -\frac{3}{2}\theta_2(t)\right)+\sin\left(\sqrt{3}\theta_1(t)\right)\biggl\}.
    \end{array}
\label{eq:phasemodh}
\end{equation}
For the brick-wall Haldane model, the amplitude modulation is simpler, given by
\begin{eqnarray}
    \small
    \begin{array}{ll}
    V_x(t) &= V_0 \biggl\{2\cos\left(\theta_1(t)\right) + \cos\left(\theta_2(t)\right)\biggl\} \\
    
    V_y(t) &= -V_0 \biggl\{\sin\left(\theta_2(t)\right)\biggl\}
\end{array}
\label{eq:ampmodbwh}
\end{eqnarray}
and the frequency modulation is given by
\begin{equation}
    \lambda(t) = -2gV_0\sin(\phi)\bigl\{ \,\text{sin}\left(\theta_2(t) - \theta_1(t)\right) - \text{sin}\left(\theta_2(t) + \theta_1(t)\right)\bigl\}.
    \label{eq:phasemodbwh}
\end{equation}
We choose frequency scales for $\Omega_1$ and $\Omega_2$ so as to evolve the system adiabatically, i.e., $\Omega_1$, $\Omega_2 \ll gV_0$, with $ gV_0 \ll \mu$ coming from the rotating wave approximation of the interaction picture. The RF detuning $\delta$ in Eq. \eqref{eq:floquetHamiltonian} now embodies the crucial mass term $M$ that can be varied to traverse the topological phase diagram of the Haldane Hamiltonian. For all results henceforth, we assume $\Omega_R = gV_0/2$ and $M = \delta/\Omega_R$. 

Adding an external optical drive and photon loss to our resonators, we get the supermode evolution equations,
\begin{equation}
    \dot{b}_j = i\left[H, b_j\right] - \gamma b_j + \sqrt{\gamma_e}\text{s}_{\text{in}}(t)e^{\pm i((\mu + \delta)t+\Delta(t))/2}
\end{equation}
where $\gamma_e$ is the coupling rate into  the bus-waveguide, $\gamma$ is the loss associated with the resonators and $\text{s}_{\text{in}}(t)$ is the external drive. Here, $j \in \{1, 2\}$. The $+,-$ sign in the last exponent corresponds to $j=1,2$ respectively.
 
The work done by each drive is then calculated using the equation,
\begin{equation}
    W_i = \int_0^T dt' \Omega_i\left< \frac{\partial \mathcal{H}(t')}{d\theta_i} \right>; \ \ \ i=\{1,2\}
    \label{eq:workFormula}
\end{equation}
Previous work in studying quantized pumping using Floquet synthetic dimensions \cite{SashankPaper, MRH, malz_topological_2021} was based on a square lattice with NN hopping, such as the Qi-Wu-Zhang model. 
\begin{eqnarray}
    \nonumber \mathcal{H}_{\frac{1}{2}\text{BHZ}} & = &\Omega_R[\sin(\theta_1(t))\sigma_x  + \sin(\theta_2(t))\sigma_y + \{M - \cos(\theta_1(t)) - \cos(\theta_2(t))\}\sigma_z]\\
    &=& h_1(t) + h_2(t) + M\Omega_R\sigma_z
\end{eqnarray}
In the above equation, $h_1(t) = \Omega_R\left(\text{sin}(\theta_1(t))\sigma_x -\text{cos}(\theta_1(t))\sigma_z\right)$ and contains contributions only from the drive with frequency `$\Omega_1$', while $h_2(t) = \Omega_R\left(\text{sin}(\theta_2(t))\sigma_y - \text{cos}(\theta_2(t))\sigma_z\right)$ and contains contributions only from the drive with frequency `$\Omega_2$'.
The separability of the drives along orthogonal directions in these models makes it straightforward to calculate the work done.
\begin{equation}
     W_i = \int_0^T dt' \left< \frac{\partial h_i(t')}{dt'} \right>.
    \label{eq:workFormula}
\end{equation}
This does not hold true in general, however, as the contributions from each drive in the Floquet Hamiltonians of the Haldane and brick-wall Haldane models are inseparable due to the non-orthogonality of lattice vectors in such non-square lattices. We refer to the brick-wall Haldane as a non-square lattice too because of the presence of $45$\textdegree $\,$diagonal NNN couplings and its topological similarity to the honeycomb-lattice Haldane model. In either case, the Berry curvature leads to quantized energy pumping and the Floquet lattice transport is characterized by the work extracted from the drives:
\begin{equation}
    \frac{1}{2}\frac{\partial}{\partial t}(W_2-W_1)  = \mathcal{C}\frac{\Omega_1\Omega_2}{2\pi} \label{eq:quantized_pumping}
\end{equation}
We now turn to numerical simulations to confirm this claim.
\begin{figure*} 
    \hspace{-.8in}
   \includegraphics[scale=0.25]{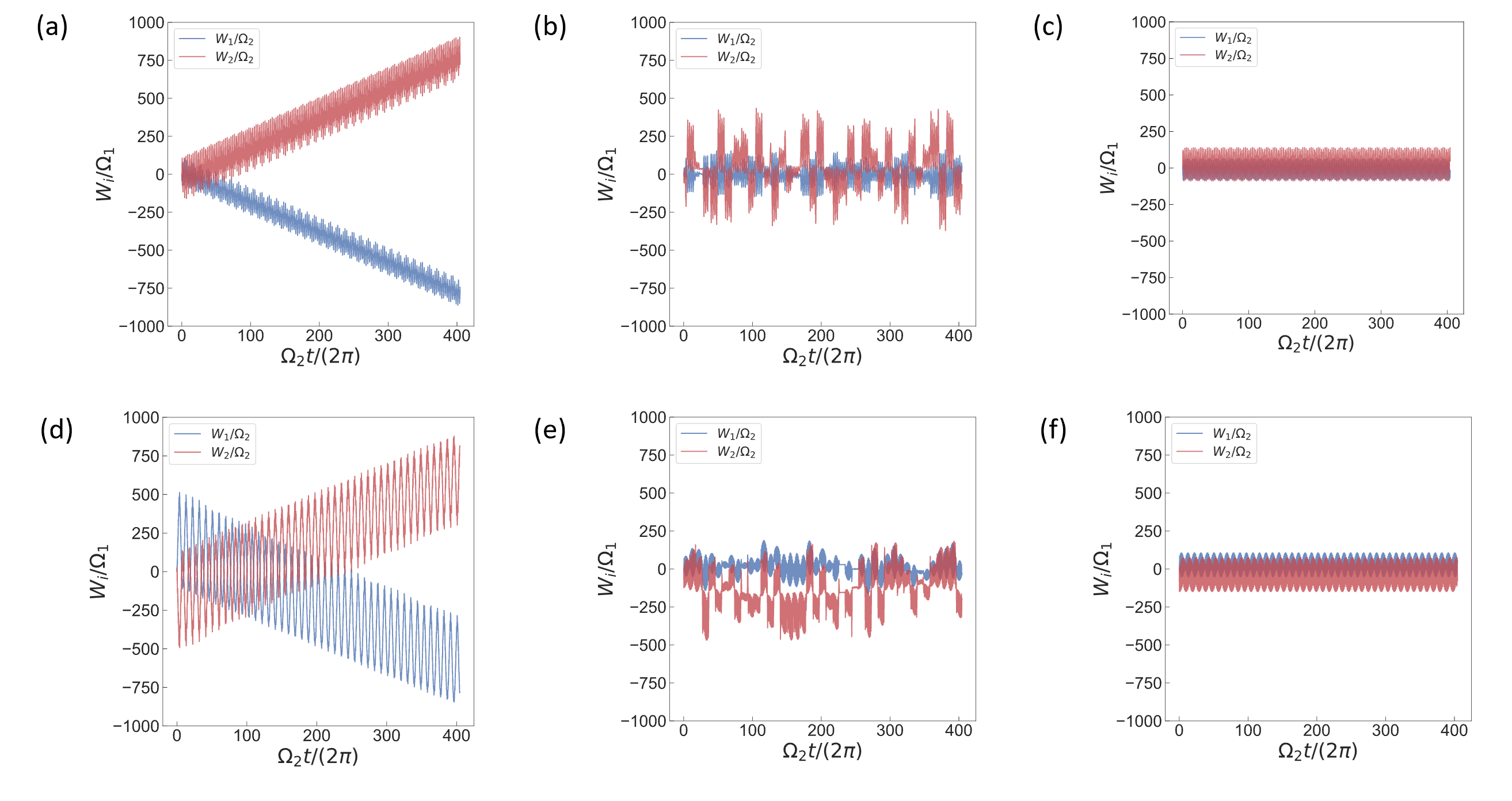}
    \caption{Work done between incommensurate drives of the brick-wall Haldane/Haldane models (a)/(d) Topological phase $M = 1$, (b)/(e) Phase boundary $M = 2\sqrt{3}\sin(\phi)$/$M = 3\sqrt{3}\sin(\phi)$ and (c)/(f) Trivial phase $M = 6$. Parameters used for simulation are $\Omega_2/\Omega_1 = (1 + \sqrt{5})/2$ with $\Omega_1 = 3$ MHz, $\Omega_R = 125$ MHz, $\mu$ = 30 GHz, $\gamma = \gamma_e = 0$, $s_{\text{in}}(t) = 0$, $\phi_1 = \pi/10$, $\phi_2 = 0$ and $\phi = \pi/2$. For the topological phase, we see quantized pumping between drives of incommensurate frequencies, as evidenced by slopes of work plots in (a) of $\pm 1.97$, and slopes in (d) of 1.96 and -1.94. At the phase boundary, the transition between topological and trivial phases become evident: we stop seeing quantized pumping along with the work becoming jagged with sharp jumps. The trivial phase shows no energy transfer between the pumps.}
    \label{fig:HamiltonianDyn}
\end{figure*}



\begin{figure}
    \centering
    \includegraphics[scale=0.25]{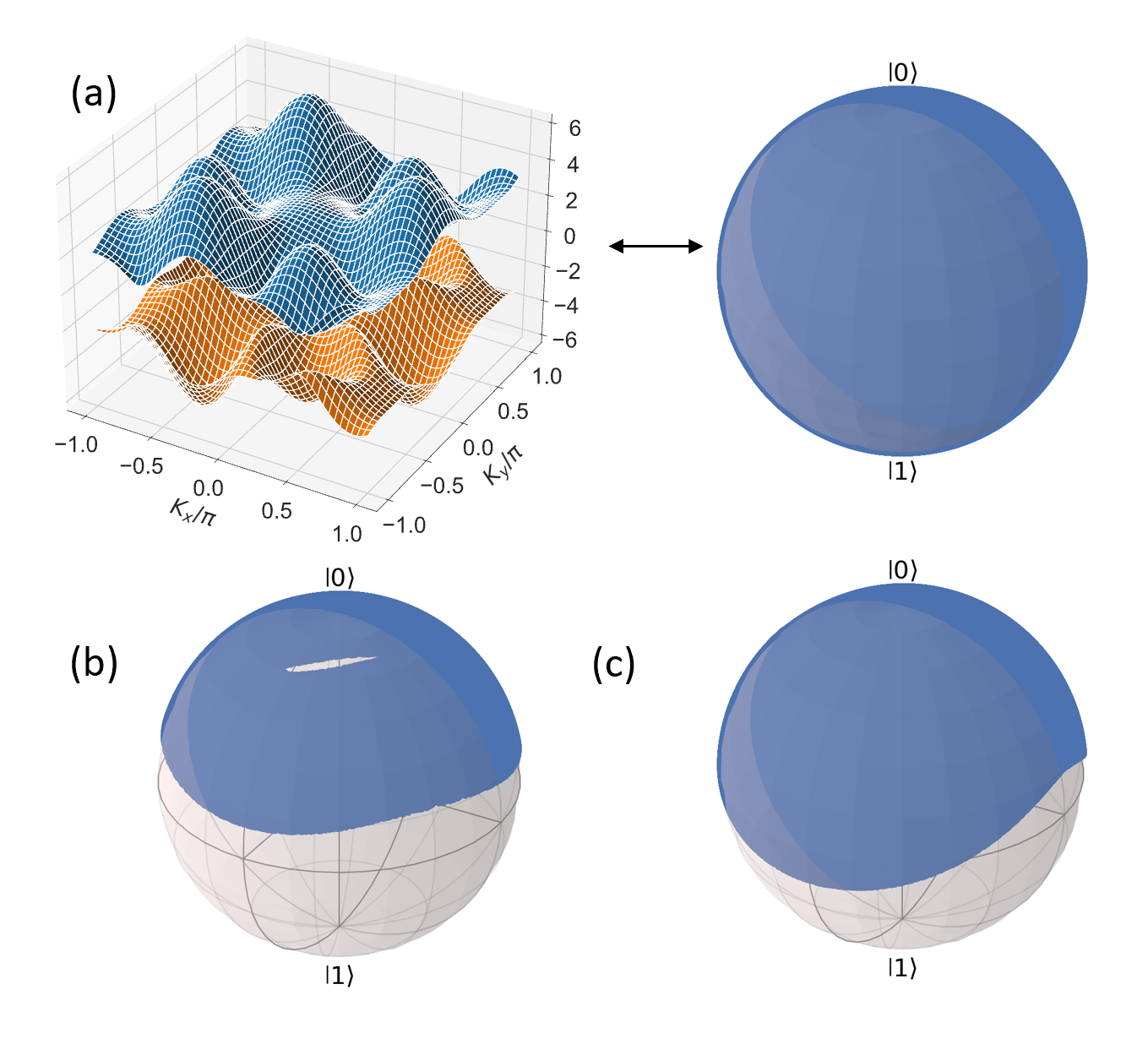}
    \caption{(a) Complete mapping of the Brillouin zone (torus in this case) to the Bloch-sphere. In the topological phase, the incommensurate nature of the drives forces no repetition of the dynamics, which leads to the complete mapping of the Bloch-sphere. We also observe complete mapping at the phase boundary. Trivial phase of the Haldane (b) and brick-wall Haldane (c) shows localization of the points.}
    \label{fig:BlochSpherePhaseSpaceDisc}
\end{figure}
\begin{figure}[h]
    \includegraphics[scale=0.32]{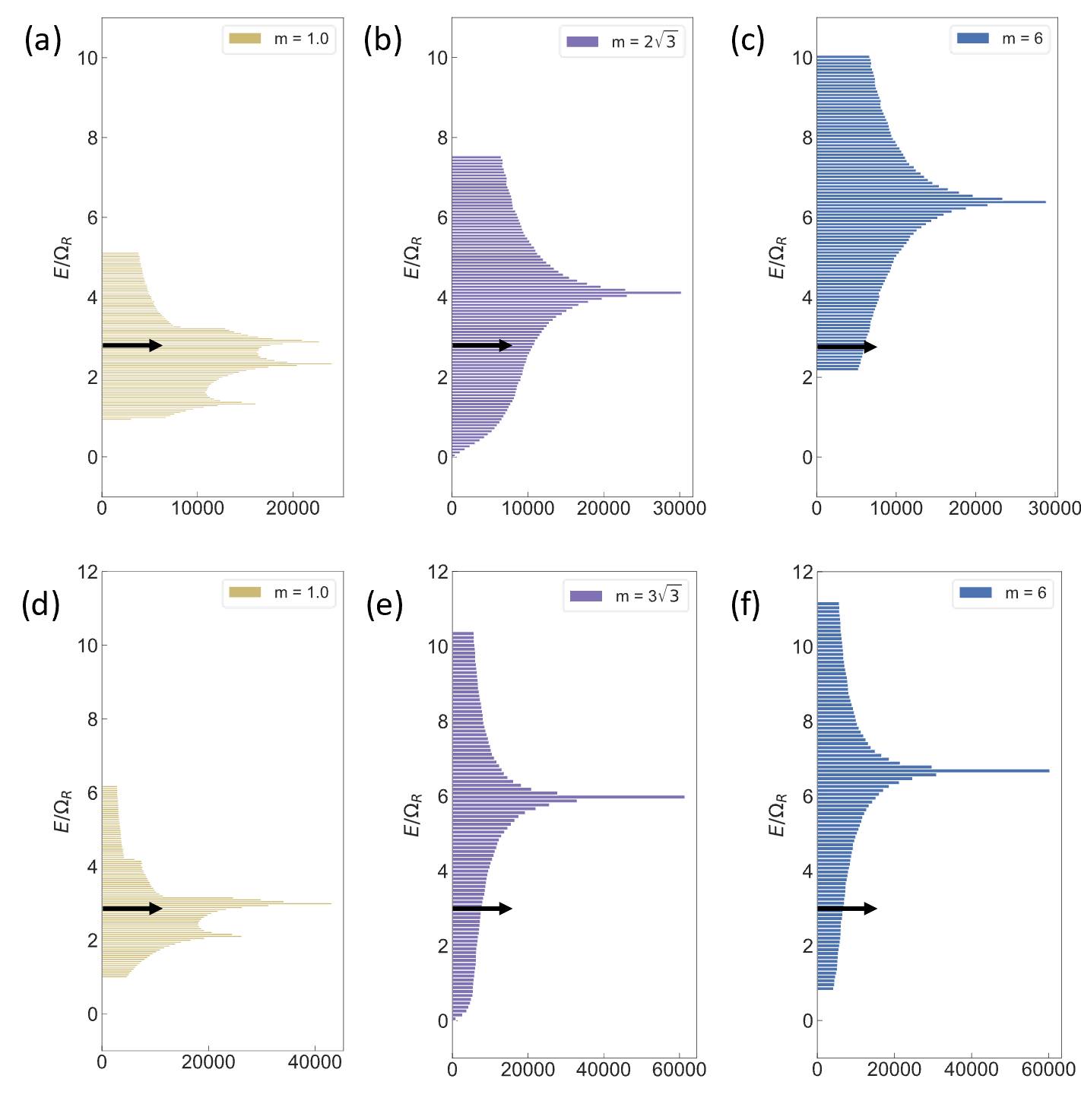}
    \caption{(a)-(c): Density of States (DoS) for the brick-wall Haldane Model in the topological phase, phase boundary and trivial phase, respectively. (d)-(f): Density of States for the Haldane model in the same phase as above. To motivate the chosen detuning of the laser, we look at the DoS for the case of no drive and dissipation. When the frequency of the laser is chosen as $\omega_D = \omega_0 - \mu/2 - 3\Omega_R$, it resonantly drives the supermodes at the instant $E = 3\Omega_R$. Note that we can rotate out the $\omega_0$ and $\mu$, so the non-zero DoS at $E = 3\Omega_R$ motivates this frequency. Small differences in the DoS are caused due to the contrasting structure of BZ ($\pi/4$ rotated square for the brick-wall Haldane and regular hexagon for Haldane model)}
    \label{fig:All_DOS}
\end{figure}

\subsection{\label{sec:level6a}Analysis of conservative Hamiltonian dynamics}

Before understanding the effect of drive and dissipation on the topological phase, it is necessary to understand the conservative evolution of the states according to the topological Hamiltonian. For this case, we set $\gamma = \gamma_e = 0$ and $\text{s}_{\text{in}} = 0$, thus turning off the drive and dissipation. Here, initializing the supermodes to be a Floquet eigenstate of the initial Hamiltonian at time $t=0$ provides a good understanding of the dynamics.

Our numerical analysis shows that the work done by the drives in the topological phase is quantized. In the topological phase, that is $M=1$, for the brick-wall Haldane (Fig. \ref{fig:HamiltonianDyn}(a)) and Haldane models (Fig. \ref{fig:HamiltonianDyn}(d)), we see that the pumping in one of the drives increases linearly with time while it decreases for the other. However, we find that due to the presence of long-range couplings beyond nearest neighbor and due to the geometry of the non-square lattice, the slope of the quantized pumping is changed to twice the Chern number: $W_{1,2} \approx \pm 1.97$ for the brick-wall Haldane model and $1.96, -1.94$ for the Haldane model. In section \ref{sec:level6c} we provide a geometric argument to explain the physical origin of the doubling that is observed in the quantized transport. At the phase boundaries $M= 2\sqrt{3}\text{sin}(\phi)$ for the brick-wall Haldane (Fig. \ref{fig:HamiltonianDyn}(b)) and $M=3\sqrt{3}\text{sin}(\phi)$ for the Haldane (Fig. \ref{fig:HamiltonianDyn}(e)) models, the pumping between the drives exhibits sharp, somewhat random jumps, and this jagged pumping indicates the transition from the topological to the trivial phase. In the trivial phase at $M=6$, for the brick-wall Haldane (Fig. \ref{fig:HamiltonianDyn}(c)) and Haldane (Fig. \ref{fig:HamiltonianDyn}(f)) models, we observe no pumping between the drives. 


The presence of the topological phase is also evident from the trajectories of the spin representation on the Bloch sphere (Fig. \ref{fig:BlochSpherePhaseSpaceDisc}(a) - \ref{fig:BlochSpherePhaseSpaceDisc}(c)). In the topological phase (including the phase-boundary) we see complete coverage, informing us that every point on the {Brillouin zone} is being mapped to a sphere ($\mathbb{T}^2 \rightarrow \mathbb{S}^2$). The complete mapping in the topological phase is due to the incommensurate nature of the drives. Under this condition, the system is ergodic, leading to complete coverage in the steady state. In the trivial phase, the spin stays localized near the initial excitation (symmetric supermode) and the surface of the Bloch sphere is not completely covered. 



\subsection{\label{sec:level6b}Steady-State analysis of the Driven-Dissipative dynamics}

When adding an external drive, it is essential to first decide its detuning from the supermodes' resonance frequencies. To do so, we plot the density of states (DoS) in the non-driven case. Consider the time varying Hamiltonian, given by Eq. (\ref{eq:floquetHamiltonian}), as a function of $M$. The 2D DoS informs us how often the eigenstates are being populated by the laser at resonance. When the laser frequency $\omega_D := \omega_0 - \mu/2 - 3\Omega_R$, with $\omega_0$ being the single-ring resonance frequency and $\mu$ the separation between the supermodes, the system is resonantly driven for some fraction of the time when the eigenenergy $E(t)$ is in the vicinity of $E = 3\Omega_R$. Note that in our simulations we rotate out $\omega_0$ and $\mu$. From Fig. (\ref{fig:All_DOS}), the DoS at $3\Omega_R$ is non-zero for the entire range of $M \in [1,6]$ during the transition between trivial and topological phases, which motivates the chosen detuning of the laser $\omega_D$.
\begin{figure*}
\hspace{-0.8cm}
    \hspace{-0.8cm}
    \includegraphics[scale=0.32]{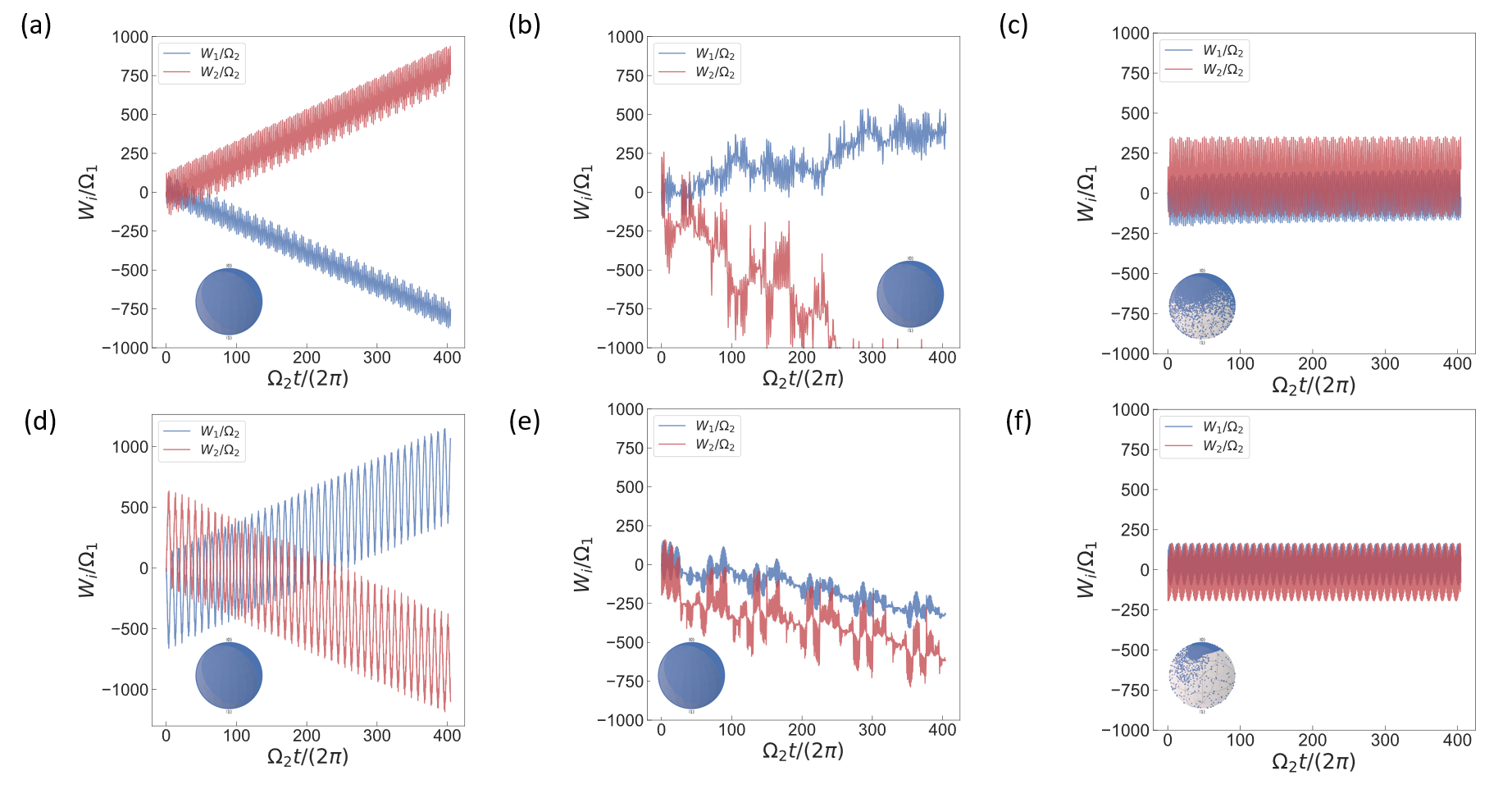}
    \caption{Same as in Fig. \ref{fig:HamiltonianDyn}, except with the addition of drive and dissipation. Insets show coverage of the Bloch-sphere in that particular phase. Parameters used for simulation are $M = 1$ (topological phase) $M = 6$ (trivial phase), $\Omega_2/\Omega_1 = (1 + \sqrt{5})/2$ with $\Omega_1 \simeq 3$ MHz, $\Omega_R = 125$ MHz, $\mu$ = 30 GHz, $\gamma = \gamma_e = 0.01$ MHz, $\phi_1 = \pi/10$, $\phi_2 = 0$ and $\phi = \pi/2$. For the brick-wall Haldane model, one can see in the topological phase, quantized pumping with slope 2.00 and -2.02. For the Haldane model, we see quantized pumping in the topological phase with slope $\pm2.60$. In this phase, both the models also show full coverage of the Bloch-sphere, even with the presence of drive and dissipation. At the phase boundary, one can observe the jaggedness in work done, indicating the phase transition, in addition to the complete coverage of the Bloch-sphere. In the trivial phase, the drives show no pumping and the Bloch-sphere shows incomplete coverage.}
    \label{fig:DD}
\end{figure*}
The DoS for the two models exhibits qualitative similarities, with multiple peaks in the topological regime and a single peak once $M$ crosses into trivial regime. However, quantitative differences in the DoS are seen in Fig. \ref{fig:All_DOS}.



In the presence of drive and dissipation, we observe that the topological phases of the brick-wall Haldane and Haldane models (Figs. \ref{fig:DD}(a, d)) show quantized pumping. For the brick-wall Haldane case, the slopes are found to be $2.00$ and $-2.02$, while in the Haldane case, the slopes are $\pm 2.60$, somewhat larger than the conservative case. Another discrepancy we notice is the change in signs for the work done by $W_{1(2)}$ in the Haldane model. Despite these differences, however, the phase boundary (Figs. \ref{fig:DD}(b, e)) shows the jaggedness in the work done, evidently similar to the conservative case (Fig. \ref{fig:HamiltonianDyn}(b, e)) and indicative of the phase transition. In the trivial phase (Fig. \ref{fig:DD}(c) and \ref{fig:DD}(f)), the lack of pumping between the drives persists. 

The Bloch-sphere trajectories (insets of the plots in Fig. \ref{fig:DD}) show complete coverage of the sphere in the topological phase and at the phase boundary, re-confirming the complete mapping from the torus to the sphere discussed in Fig. \ref{fig:BlochSpherePhaseSpaceDisc}. In the trivial phase, we observe incomplete mapping of the Bloch-sphere. Qualitatively, the addition of drive and dissipation does not seem to alter the work done by the drives significantly, attesting to the topological robustness offered by the Haldane models.

\begin{figure*}
    \centering
    \hspace{-1cm}
    \includegraphics[scale=0.28]{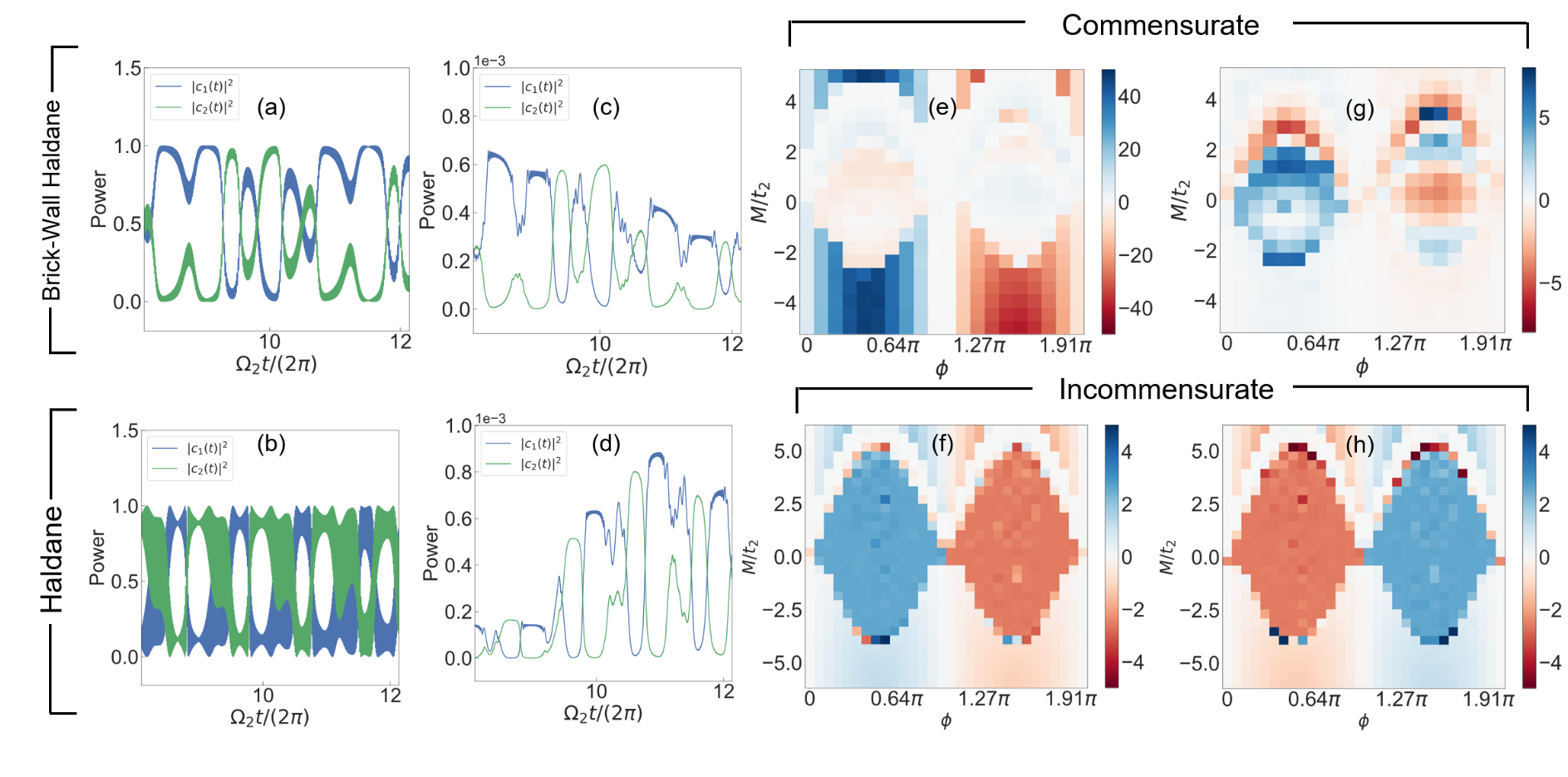}
    \caption{Supermode trajectories for the brick-wall Haldane model (a) without dissipation and (c) with drive-dissipation. Supermode trajectories for the Haldane model (b) without dissipation and (d) with drive-dissipation. Parameters used for the simulation are in the captions for Fig. \ref{fig:HamiltonianDyn} and Fig. \ref{fig:DD}. We observe the signatures of topology more or less preserved even with the addition of drive and dissipation, with reduced micromotion in the trajectories of the supermodes in (c) and (d) as compared to (a) and (b) respectively, but similar envelope dynamics when renormalized. 
    Slopes of the work done by drives 1 (e and f) and drive 2 (g and h) for a parameter sweep of $M$ and flux $\phi$.
    (e), (g) (brick-wall Haldane; Commensurate), (f) and (h) (Haldane; Commensurate) show a heatmap for the slopes at each point in the discretization of the phase space. There is excellent agreement between the well know phase plot for the Haldane model (Fig. \ref{fig:HaldBWHAllPlots}(d)), and what we observe on a Floquet lattice (Fig. \ref{fig:allmodesplot}(f, h)). The  normalized pumping rate ($2\pi \dot W_i/(\Omega_1 \Omega_2)$ is close to twice the Chern number throughout the phase diagram, and it is also equal and opposite for the two drives. Similar phase plots were seen for a change in initial conditions, i.e., $\phi_1$ and $\phi_2$, indicating ``averaging out" because of the addition of drive and dissipation. We also observe anomalously high slopes at the transition point between the trivial and topological phases as the dynamics is crossing a Dirac point. In Fig. \ref{fig:allmodesplot}(e, g), we simulate a commensurately driven brick-wall Haldane ($\Omega_1/\Omega_2=1.5$), which results in loss of the topological structure  --  the pumping between the drives is not quantized anymore. More importantly, the normalized pumping rate is not equal and opposite for the two drives. We also see other signatures like anomalously high pumping in regions corresponding to the "trivial" case of the incommensurately related frequencies ($M<-3$).}
    \label{fig:allmodesplot}
\end{figure*}

Thus, a remarkable feature of these models is that the signatures of temporal topology, i.e., work done, Bloch-sphere coverage etc. (Fig \ref{fig:DD}) remain preserved with the addition of drive and dissipation. This is visualised in Fig. \ref{fig:allmodesplot}, in the trajectories of the supermodes in the brick-wall Haldane (Figures \ref{fig:allmodesplot}(a, c) for Hamiltonian and driven-dissipative dynamics respectively) and Haldane (Figures \ref{fig:allmodesplot}(b, d) for Hamiltonian and driven-dissipative dynamics respectively) models. Parameters used for this simulation are the same, given in the captions of Figs. \ref{fig:HamiltonianDyn} and \ref{fig:DD}. Although the trajectories of the supermodes for a given model are different between the conservative and the driven-dissipative evolution, the adiabatically-varying envelopes appear quite similar within the renormalized subspace, indicating that the topology is well-preserved beyond the photon lifetimes due to the continuous repopulation of the state space. Moreover, the presence of drive and dissipation is observed to reduce the micromotion, which has a stabilizing effect on the topological dynamics \cite{ritter_autonomous_2024}. This can be intuitively interpreted to arise from the nontrivial topology - behavior that is locally different (such as the microscopic time evolution) but shows globally similar properties (such as rate of work done). 

To fully realize the Haldane model phase diagram on Floquet synthetic dimensions with drive and dissipation, we can discretize the phase space given in Fig. \ref{fig:HaldBWHAllPlots}(d) to solve for the slopes of the pumping. Figs. \ref{fig:allmodesplot}(f, h) show the slopes for drive one and drive two for a $30\times30$ discretization of the phase space. Parameters used for the simulation are the same from Fig. \ref{fig:HamiltonianDyn} and \ref{fig:DD}. We observe good agreement between the phase plots of the Haldane model and the rate of energy pumping in the  Floquet-synthetic-dimension encoded Haldane model. We also observe similar trends with close to twice the Chern number quantized energy pumping. The drives show approximately equal and opposite slope for each point in the phase space. It is to note that the phase space of the Floquet Haldane model under the effects of drive and dissipation is insensitive to initial conditions, i.e., by changing the values of $\phi_1$ and $\phi_2$ at time $t=0$, the steady state remains unchanged. This happens because of the ``averaging out'' of the dynamics caused by the addition of the drive and dissipation. We also observe anomalously high slope values on the edge of the trivial and topological phases, which happens due to the dynamics crossing a Dirac point where the band gap closes and the trajectories cross the high-Berry-curvature region.

For the brick-wall Haldane model (Fig. \ref{fig:allmodesplot}(e, g)) we run a similar algorithm for a $20\times20$ discretization, but with the frequencies being commensurately related. Primarily, we observe that pumping between the drives is not quantized anymore, and the rates of pumping are not equal and opposite to each other. In the region corresponding to the "trivial" regime of the incommensurately related case, we observe that the pumping is anomalously high, while in other regions we also observe pumping with the same slope. Such signatures indicate the absence of topological structure. 
We now analyze the discrepancy of the Chern number that we observe in numerical simulations, as compared to our calculations.

\subsection{\label{sec:level6c} Geometric argument for twice-Chern number pumping}

In Figs. \ref{fig:HamiltonianDyn} and \ref{fig:DD}, we observed quantitative deviations in the rate of energy pumping, which consistently exceed the expectation in Eq. \eqref{eq:quantized_pumping} based on the Chern number by a factor of 2 or more. This seems to indicate unique behaviour in the Haldane model, independent of the lattice geometry it is realized in - full Haldane or the simplified brick-wall Haldane. Here we provide geometric argument for this increase in the pumping rate.
In contrast to previous work in Floquet synthetic dimensions, which only considered square lattices with NN hopping, our analytical calculations and numerical simulations suggest that the NNN hopping terms provide an additional pathway to the anomalous velocity, thereby contributing to the work done. 

For the brick-wall Haldane model, in addition to the $\theta_1(t)$ and $\theta_2(t)$ drives, to probe the non-square nature, we would need extra tones of frequency $\theta_2(t) \pm \theta_1(t)$, giving us the net rate of energy transfer: 
    \begin{align}
    \nonumber
    \text{P} &= \frac{1}{2}\frac{\partial}{\partial t}\left(\text{E}_{\Omega_1} - \text{E}_{\Omega_2}\right) + \frac{1}{2}\frac{\partial}{\partial t}\left(\text{E}_{\Omega_2 + \Omega_1} - \text{E}_{\Omega_2 - \Omega_1}\right)\\ 
    &= \frac{1}{2}\bigl\{(\Omega_1, -\Omega_2) + (\Omega_1, \Omega_2) - (-\Omega_1, \Omega_2)\bigl\} \, \, \cdot \, \, (\vec{\Omega} \cross \mathcal{F}_{\vec{\theta}})\\ \nonumber
    &= 2\frac{C}{2\pi} \Omega_1 \Omega_2
    \end{align} 
    \label{eq:workbwh}
In the above equations, $E_{a\Omega_1 + b\Omega_2}$, where $a, b \in \mathbb{R}$, is the instantaneous energy due to the drive $(a\Omega_1, b\Omega_2)$ being applied on the Floquet lattice. 

Equivalently, for both the models, the net transfer of photons should be such that it is in the direction of $\left(-\Omega_2, \Omega_1\right)$, so the response can be decomposed into an $x$-component and a $y$-component. Owing to the non-square nature of the Floquet lattices, the paths involving NN and NNN hopping average out to be in this direction.

For the Haldane model, this is achieved by two NN hopping ($\vec{a}_1$ followed by $-\vec{a}_2$) or one NNN hopping ($\vec{b}_3$) and four NN hopping ($-\vec{a}_2$, $\vec{a}_3$, $-\vec{a}_1$ and $\vec{a}_3$) or 2 NNN hopping ($\vec{b}_1$ and $\vec{b}_2$) in the $y$-direction. Mathematically, the net power in the $x$-direction is given by:

\begin{equation}
    \text{P}_{x} = \frac{1}{2\sqrt{3}} \frac{\partial}{\partial t}\biggl\{\left[\text{E}_{\vec{a}_1 \cdot \vec{\Omega}} - \text{E}_{\vec{a}_2 \cdot \vec{\Omega}}\right] + \text{E}_{\vec{b}_3 \cdot \vec{\Omega}}\biggl\} = \frac{C}{2\pi}\Omega_1 \Omega_2
    \label{eq:workxhald}
\end{equation}
where the contributions due to the first two terms and minus the third term are same. Physically, this means that the path the photon takes is an average of the two. In addition, we also normalize the power to the length of the net hopping, in order to scale the two directions to equal footing. 

For the net power in the $y$-direction, 
\begin{align}
    \text{P}_y &= \frac{1}{2}\frac{1}{3} \frac{\partial}{\partial t}\biggl\{\left[-\text{E}_{\vec{a}_2 \cdot \vec{\Omega}} + \text{E}_{\vec{a}_3 \cdot \vec{\Omega}} - \text{E}_{\vec{a}_1 \cdot \vec{\Omega}} + \text{E}_{\vec{a}_3 \cdot \vec{\Omega}}\right]  \nonumber \\ \ \ &\ \ -
    \text{E}_{\vec{b}_1 \cdot \vec{\Omega}} - \text{E}_{\vec{b}_2 \cdot \vec{\Omega}}\biggl\} = \frac{C}{2\pi}\Omega_1 \Omega_2.
    \label{eq:workyhald}
\end{align}
Hence, the total power exchanged in the Haldane model is given by
\begin{equation}
    \text{P} = \text{P}_x + \text{P}_y = 2\frac{C}{2\pi} \Omega_1 \Omega_2.
\end{equation}

\section{\label{sec:level7}Discussion and Experimental Prospects}
In this paper, we have proposed and numerically demonstrated that quantized topological pumping in non-square lattice Hamiltonians such as the Haldane model can be observed in a dynamically-modulated photonic molecule, even in the presence of optical drive and dissipation. 
Our proposal uses Floquet synthetic dimensions and allows for Hamiltonian simulation of arbitrary lattice geometries directly in $k$-space without requiring modifications or simplifications of the lattice that were needed with previous photonic synthetic dimension platforms. 
A striking feature of the driven-dissipative photonic molecule is that despite the optical drive qualitatively altering the spin's trajectory, in comparison to the Hamiltonian case, the topological structure of the model is more or less preserved.

To achieve this photonic molecule in practice, we envision an electro-optically modulated high-Q coupled-ring resonator device, that can be achieved with state-of-the-art integrated lithium niobate platforms \cite{zhang_electronically_2019, gao_lithium_2022}. With all the frequency scales of our setup being solely upper bounded by $\mu$, which is set by the coupling gap between the rings, we can circumvent the bandwidth limitations of on-chip ring modulators and conventional photonic frequency synthetic dimension platforms.
However, the added complexity of lattice geometry and long-range couplings is transformed into more demanding RF-signal generation, as additional harmonics of the incommensurate drives are required to achieve the required model. This is nevertheless within the capabilities of high-speed direct digital synthesizers and arbitrary waveform generators (AWGs).

Looking ahead, photonic Floquet lattices hold substantial promise for simulating general 2D and higher-dimensional topological models. By increasing the real-space complexity of our devices, or coupling to other synthetic dimensions, we envision pathways to achieving higher-dimensional lattices with well-defined boundaries beyond the traditional square, cubic and hypercubic geometries. This, in turn, has the potential to facilitate a new class of topologically-robust photonic devices and novel modalities of optical frequency control.

\section{\label{sec:level9}Acknowledgments}
The authors acknowledge the University of Maryland super-computing resources (\href{http://hpcc.umd.edu}{http://hpcc.umd.edu}) made available for conducting the research reported in this paper. The authors thank Antoine Henry for carefully reading this manuscript, as well as Anushya Chandran and David Long for stimulating discussions. S.S. thanks Jay Sau for his notes on quantum many-body theory. This work was supported by an NSF CAREER award (grant no. 2340835), an NSF QuSeC-TAQS grant (no. 2326792), and a Northrop Grumman seed grant.

\bibliography{sample, My_Library}

\end{document}